\documentclass{aa}  

\usepackage{graphicx}
\usepackage{txfonts}
\begin{document} 

\title{The broad-line region and dust torus size of the Seyfert 1 galaxy PGC50427\thanks{Based on observations made with the Southern African Large Telescope (SALT)}.\\}	
	\titlerunning{The BLR and dust torus size of PGC50427}	

	\author{
          F. Pozo Nu\~nez 
          \inst{1} 
          \and
          M. Ramolla
          \inst{1}
          \and
          C. Westhues          
          \inst{1}
          \and
          M. Haas
          \inst{1}
          \and
          R. Chini
          \inst{1,2}
          \and
          K. Steenbrugge
          \inst{2,4}		
          \and
          A. Barr Dom\'{i}nguez
          \inst{1,2}
          \and    
          L. Kaderhandt
          \inst{1}
          \and
          M. Hackstein
          \inst{1}
          \and
          W. Kollatschny
          \inst{3}
          \and
          M. Zetzl
          \inst{3}
          \and
          Klaus W. Hodapp\inst{5}
          \and 
          M. Murphy
          \inst{6}		
	}
	\institute{
          Astronomisches Institut, Ruhr--Universit\"at Bochum,
	  Universit\"atsstra{\ss}e 150, 44801 Bochum, Germany
	  \and
          Instituto de Astronomia, Universidad Cat\'{o}lica del
          Norte, Avenida Angamos 0610, Casilla
          1280 Antofagasta, Chile
        \and
          Institut f\"ur Astrophysik, 
          Universit\"at G\"ottingen, 
          Friedrich-Hund Platz 1,
          37077 G\"ottingen, Germany
          \and
          Department of Physics, 
          University of Oxford, 
          Keble Road,
          Oxford OX1 3RH, UK
          \and
  Institute for Astronomy, University of Hawaii, 640 North Aohoku Place, Hilo, HI 96720, USA
          \and 
Departamento de F\'{i}sica, Universidad Cat\'{o}lica del
  Norte, Avenida Angamos 0610, Casilla
  1280 Antofagasta, Chile  
        }
        
	\authorrunning{F. Pozo Nu\~nez et al.}

	\date{Received ; accepted}
        
	\abstract{We present the results of a three years monitoring campaigns of the $z = 0.024$ type-1 active galactic nucleus (AGN) PGC50427. Using robotic telescopes of the Universit\"atssternwarte Bochum near Cerro Armazones in Chile, we monitored PGC50427 in the optical and near-infrared (NIR). Through the use of Photometric Reverberation Mapping with broad and narrow band filters, we determine the size of the broad-line emitting region by measuring 
the time delay between the variability of the continuum and the H$\alpha$ emission line. The H$\alpha$ emission line responds to blue continuum variations with an average rest frame lag of 
$19.0 \pm 1.23$ days. Using single epoch spectroscopy obtained with the Southern African Large Telescope (SALT) we determined a broad-line H$\alpha$ velocity width of 1020 km s$^{-1}$ and in combination with the rest frame lag and adoption a geometric scaling factor $f = 5.5$, we calculate a black hole mass of $M_{BH} \sim 17 \times 10^{6} M_{\odot}$. Using the flux variation gradient method, we separate the host galaxy contribution from that of the AGN to calculate the rest frame 5100\AA~ luminosity at the time of our
monitoring campaign. We measured small luminosity variations in the AGN ($\sim 10\%$) accross the 3 years of the monitoring campaign.
The rest frame lag and the host-subtracted luminosity permit us to derive the position of PGC50427 in the BLR size -- AGN luminosity diagram, which is 
remarkably close to the theoretically expected relation of $R \propto L^{0.5}$. The simultaneous optical and NIR ($J$ and $K_{s}$) observations allow us to determine the size of the dust torus through 
the use of dust reverberation mapping method. We find that the hot dust emission ($\sim 1800K$) lags the optical variations with an average rest frame lag of $46.2 \pm 2.60$ days. The dust reverberation radius and the nuclear NIR luminosity permit us to derive the position of PGC50427 on the known $\tau - M{V}$ diagram. The simultaneus observations for the broad-line region and dust thermal emission demonstrate that the innermost dust torus is located outside the BLR in PGC50427, supporting the unified scheme for AGNs.
	}
        
	\keywords{ galaxies: active --galaxies: Seyfert --quasars: emission lines
          --galaxies: distances and redshifts --galaxies: individual: PGC50427 }
		\maketitle
%

\section{Introduction}

\begin{figure*}
  \centering
  \includegraphics[width=12cm,clip=true]{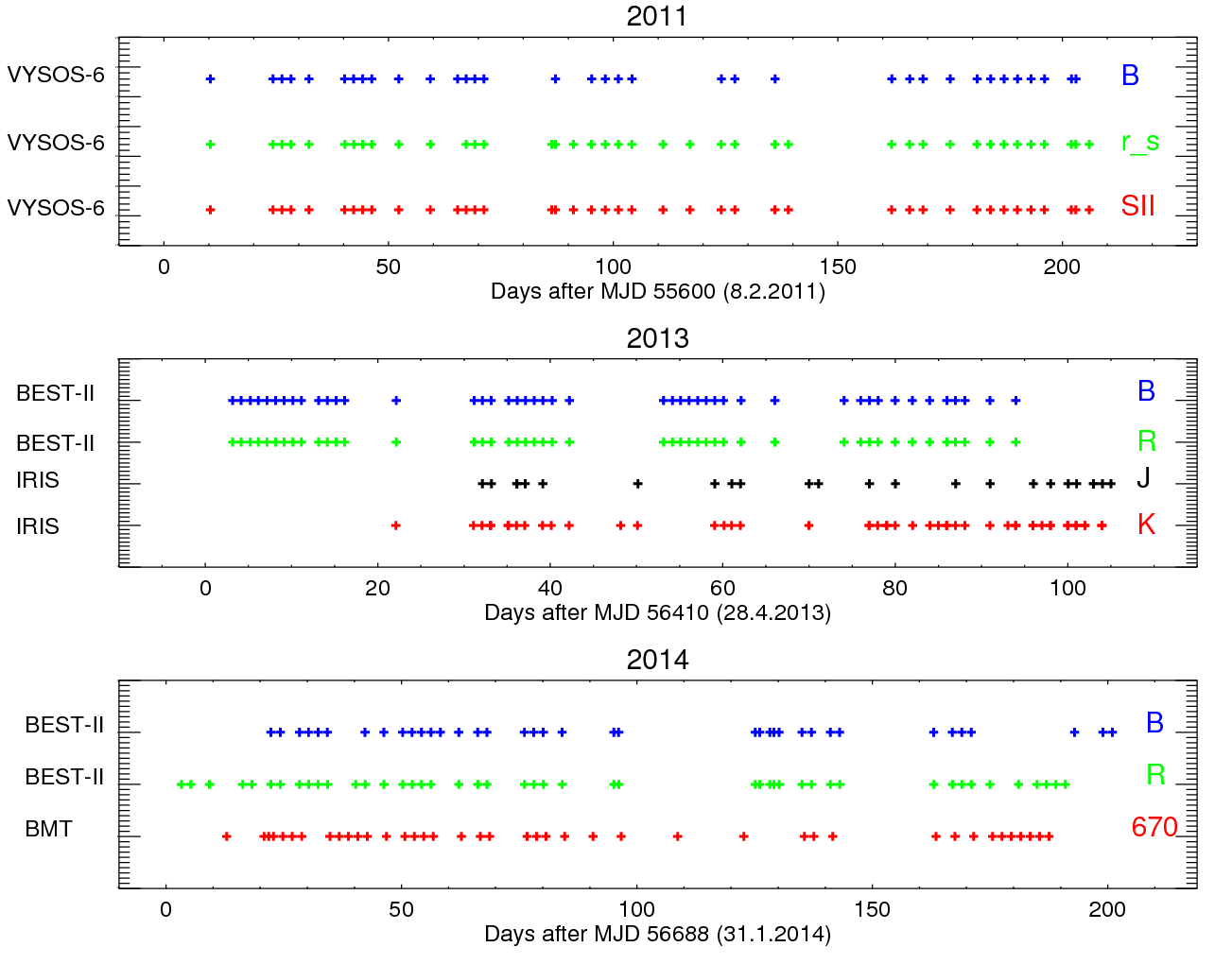}
  \caption{The timeline of our campaign on PGC50427. The days in which the telescopes made observations are indicated by crosses.}
  \label{timeline}
\end{figure*}

\begin{table*}
\begin{center}
\caption{Characteristics of \object{PGC50427}}
\label{table1}
\begin{tabular}{@{}cccccccccc}
\hline\hline
$\alpha$ (2000)$^{(1)}$ & $\delta$ (2000)$^{(1)}$ & z$^{(1)}$ & $D_L^{(1)}$ & $B-V^{(2)}$ & $M_{abs}^{(2)}$ & $A_u^{(3)}$ & 
$A_B^{(3)}$ & $A_R^{(3)}$ & $A_r^{(3)}$\\
      & & & (Mpc) & (mag) & (mag) & (mag) & (mag) & (mag) & (mag)  \\
\hline
14:08:06.7 & -30:23:53.7 & 0.024 & 102.0 & 0.40 & -20.6 & 0.252 & 0.215 & 0.129 & 0.135 \\
\hline
\end{tabular}
\end{center}
\tablefoottext{1}{Values from NED database,} 
\tablefoottext{2}{\cite{2010A&A...518A..10V},}
\tablefoottext{3}{\cite{2011ApJ...737..103S}.}
\end{table*}

The existence of a dusty structure with a torus-like geometry surrounding the broad-line region (BLR) and the central accreting super massive black hole (SMBH) 
play a 
fundamental role in the framework of an unified model for active galactic nuclei (AGN; \citealt{1993ARA&A..31..473A}). Its presence would explain the observed 
differences in the spectra of type 1 and 2 Seyfert galaxies together with the strong emission observed at near/mid/far infrared wavelengths 
(\citealt{2001AJ....121.1369A}; \citealt{2003A&A...402...87H}; \citealt{2004A&A...424..531H}; \citealt{2005A&A...436L...5S}). 

Evidence for the 
surrounding hot 
dust torus rests on indirectly optical spectropolarimetric observations (eg., \citealt{1985ApJ...297..621A}; \citealt{1987ApJ...320..537B}; 
\citealt{1993ApJ...404...94K}). A 
direct detection of the putative dust torus is more difficult since the internal structure of AGNs is spatially unresolved using single telescopes. IR 
long-
baseline interferometric observations have been able to determine the size of the dust torus for a few nearby AGNs (eg., \citealt{2004A&A...418L..39W}; 
\citealt{2007A&A...474..837T}; \citealt{2010ApJ...715..736P}). 

The only method available to study the origin and morphology of the BLR and the dust torus independent of the spatial resolution of the instrument is Reverberation Mapping (RM, 
\citealt{1972ATsir.688....1L}; \citealt{1973ApL....13..165C}; \citealt{1982ApJ...255..419B}; \citealt{1986ApJ...305..175G}; \citealt{1993PASP..105..247P}; 
\citealt{2004ApJ...613..682P}). In this method one measures the light travel time between the accretion disk (AD) and the BLR and/or hot dust. The hot AD produces a variable continuum emission, and this variability is observed with a time delay ($\tau_{BLR}=R_{BLR}/c$) in the studied broad emission lines of the BLR. Similarly, if the dust torus is located at some radial distance $R_{dust}$ around the hot AD it will reprocess the UV/optical radiation to thermal near-infrared (NIR) radiation 
with a characteristic time delay $\tau_{dust}=R_{dust}/c$. Reverberation mapping has revealed the size of the BLR, the black hole mass and Eddington ratios in about 50 AGNs (see \citealt{2014ApJ...782...45D} and references therein).

The torus reacts to the AD variability in a wide range of time, from days to decades (eg., 
\citealt{2004MNRAS.350.1049G}; \citealt{2006ApJ...639...46S}; \citealt{2014A&A...561L...8P}; \citealt{2014ApJ...788..159K}). From dust reverberation measurements, 
which are principally based on the cross-correlation analysis between the optical ($V$, 0.55 $\mu m$) and NIR ($K$, 2.2 $\mu m$) light curves, 
a correlation between the innermost radius of the torus and the square root of the optical luminosity has been shown ($R_{dust} \propto L^{0.5}$; 
\citealt{2004MNRAS.350.1049G}, \citealt{2004ApJ...600L..35M}, \citealt{2006ApJ...639...46S}; \citealt{2014ApJ...788..159K}). The dust is heated by the accretion disk (AD) until up to its maximum sublimation temperature ($\sim$ 1500K). The correlation and sublimation temperature are consistent with the theoretical prediction of the dust sublimation radius ($R_{sub}$, \citealt{1992ApJ...400..502B}), but is systematically 
smaller by a factor of three than the sublimation radius $R_{sub}$ predicted for graphite dust grains with a size of 0.05 $\mu m$ in radius (\citealt{1999AstL...25..483O}, \citealt{2007A&A...476..713K}). Some modified dust geometries involve BLR associated dust due
to winds/outflows from the accretion disk (\citealt{1994ApJ...434..446K},
\citealt{2006ApJ...648L.101E}, \citealt{2011A&A...525L...8C}).

Alternatively, theoretical simulations shows that an anisotropically illuminated dust torus, caused by an optically thick AD, places the inner concave region of the 
torus closer to the outer edge of the AD. Thereby increasing the response time of the torus, and thus explaining the systematic difference of the time delay with 
respect of the torus radius measured from time delay and from the sublimation temperature under an isotropically assumption (\citealt{2010ApJ...724L.183K} 
\citealt{2011ApJ...737..105K}). In the case of the Seyfert 1 galaxy WPVS48, \cite{2014A&A...561L...8P} argue that the sharp NIR echo observed is due to a geometrically 
and optically thick torus seen nearly face-on. In this scenario the observer only sees the facing rim of the torus wall, which lies closer to the observer than the 
torus equatorial plane and therefore leads to an observed foreshortened lag effect.

While great theoretical progress has been made, only a handful of observational measurements of the dust reverberation radius have been obtained during the last 
years. Moreover, as noted in \cite{2006ApJ...639...46S}, the importance of simultaneous BLR and dust torus size measurements provide an important step forward to 
test and constrain the actual paradigm of unification for AGNs. 

More recently, Photometric Reverberation Mapping (PRM) has been revisited and used as an 
efficient alternative to determine the BLR size, black hole masses and host-subtracted AGN luminosities (\citealt{2011A&A...535A..73H}; 
\citealt{2012A&A...545A..84P}, \citealt{2013A&A...552A...1P}). Through the combination of broad and narrow-band data, this method is used to measure the time
delay between the triggering continuum variations and the BLR emission line response, which has previously been isolated by the subtraction of the underlying continuum 
determined from the broad-band filter data and/or through a single spectrum contemporaneous with the campaign.

Feature-rich PRM light curves allow us to infer the basic geometry of 
the BLR, whether it is spherical or disk-like, and can thus constrain the unknown geometrical factor needed in converting the time lag and velocity width into a black hole mass (\citealt{2014A&A...568A..36P}). Furthermore, the use of broad-band data alone has been tested with 
satisfactory results (\citealt{2012ApJ...747...62C}; \citealt{2012ApJ...750L..43C}; \citealt{2012ApJ...756...73E}; \citealt{2013A&A...552A...1P}; 
\citealt{2013ApJ...769..124C}). In this method, two suitable chosen broad band filters are used to trace the continuum variations and to catch the emission line 
and continuum with the removal of the continuum performed in the cross correlation domain.

\begin{figure*}
  \centering
  \includegraphics[width=14cm,clip=true]{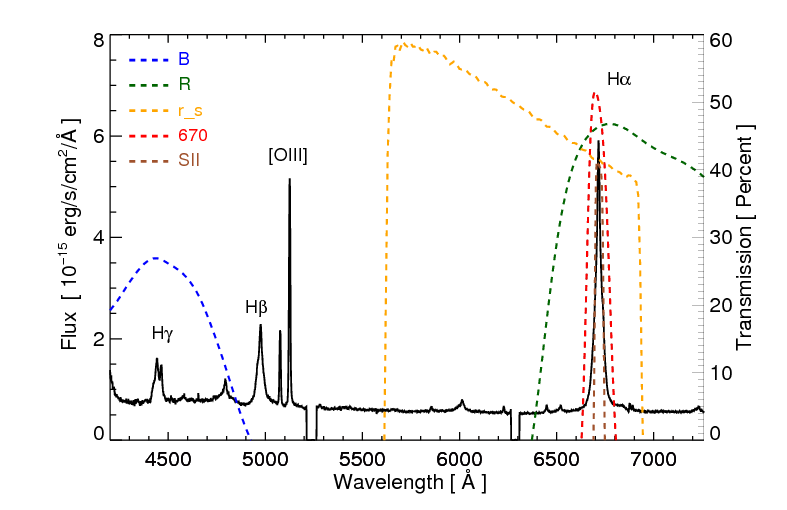}
  \caption{ 
    SALT spectrum of \object{PGC50427}. 
    For illustration, the band passes of the filters used for 
    the photometric monitoring are shown (blue $B$-band, green $R$-band, orange $r_s$-band, red $670$-band, and brown for SII-band). 
    The NB $670$ and SII catches the redshifted H$\alpha$ line, its flux is composed by the contribution of about 85\%  H$\alpha$ line and 15\% continuum. Note 
    that for actual flux calculations the filter curves are convolved with the quantum efficiency of the Alta U16 and SBIG STL CCDs cameras.}
  \label{spectrum_nbfilter}
\end{figure*}

PGC50427 has been classified as a Seyfert 1 galaxy and is located at a distance of 102 Mpc (\citealt{2010A&A...518A..10V}).
In this paper we present the results of a multi-year monitoring campaign carried out on the nucleus of the Seyfert 1 galaxy PGC50427. We use Photometric 
Reverberation Mapping in combination with dust-reverberation mapping to determine the black hole mass, the size of the BLR and dust torus for the first time in this source.

\section{Observations and data reduction}

The optical and near-infrared monitoring campaign of PGC50427 were carried out using the VYSOS-6, BEST-II, BMT and IRIS telescopes (for more details see Sections 2.1-2.2) located at the Universit\"atssternwarte Bochum observatory, near Cerro Armazones, the future location of the ESO Extreme Large Telescope (ELT) in Chile\footnote{http://www.astro.ruhr-uni-bochum.de/astro/oca/}. 

In addition to the photometric observations, one single epoch spectrum was acquired using the Robert Stobie Spectrograph (RSS) at the Southern African Large 
Telescope (SALT). The timeline of all the observations in our campaign is shown in Fig~\ref{timeline}. 

PGC50427 lies at redshift $z=0.0236$, therefore the 
H$\alpha$ emission line falls into the SII $6721 \pm 30$\,\AA and $6700\pm 60$\,\AA narrow-band (NB) filters. The characteristics of the source and the galactic foreground extinction values are listed in Table 1. Figure 2 shows the position of the NB filters with 
respect to the H$\alpha$ emission line together with the effective transmission of the other optical filters used.

\subsection{Optical monitoring} \leavevmode \\

VYSOS-6\\\\
Broad-band Johnson $B$ ($4330$\,\AA), Sloan-band $r$ ($6230$\,\AA), and narrow-band SII ($6721 \pm 30$\,\AA~, the position of the redshifted H$\alpha$ line) observations were carried out during a monitoring campaign between February 18 and September 01 of 2011, with the robotic 
VYSOS-6 telescope. The VYSOS-6 telescope consist of two 15 cm refractors (Takahashi TOA 150F Ortho-Apo-chromat Triplet) installed on a common equatorial mount 
(Bisque Paramount ME). Both refractors are equipped with an Apogee ALTA U16M $4096 \times 4096$ pixel CCD, providing a field of view (FoV) of about $2.7^{\circ} 
\times 2.7^{\circ}$. More information about the telescope and the instrument has been published by \cite{2012AN....333..706H}.

\begin{figure*}
  \centering
  \includegraphics[width=\columnwidth]{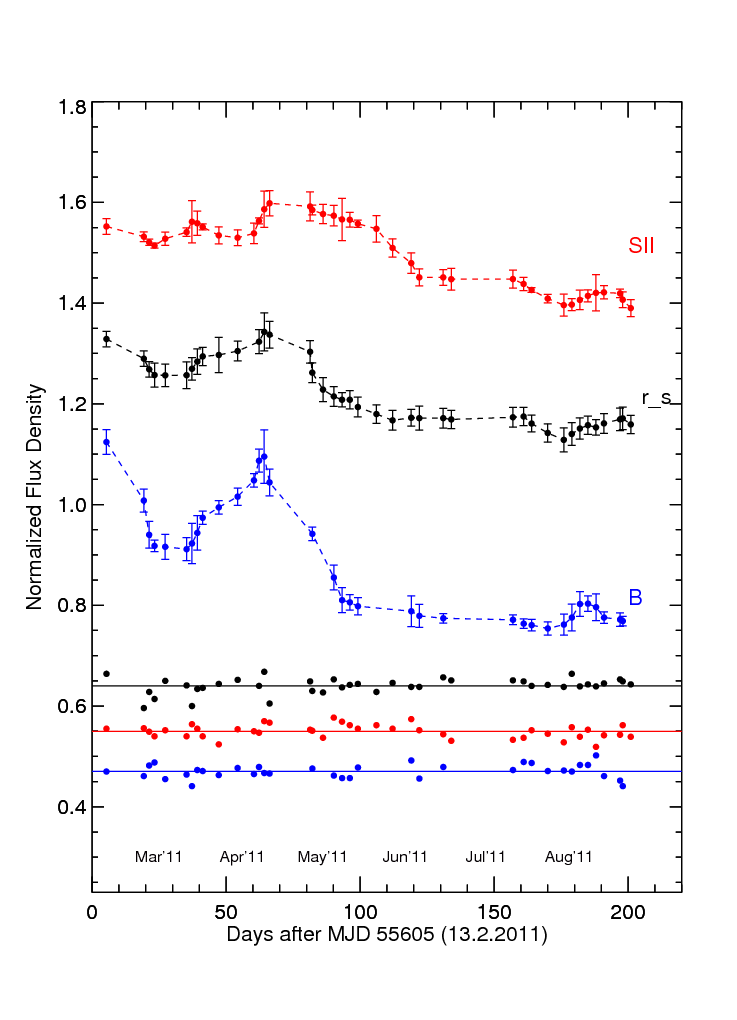}
  \includegraphics[width=\columnwidth]{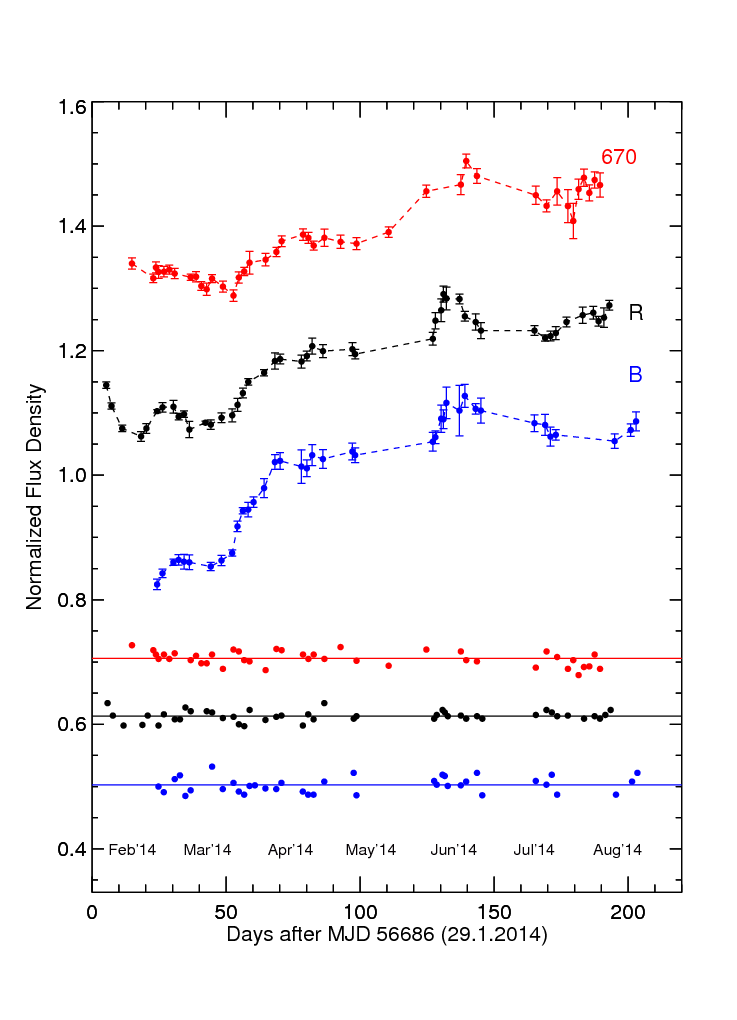}
  \caption{Observed light curves of PGC50427, as well as for some of the reference stars in the field of view for the period between February 2011 and September 2011 ({\it 
  left}) and for the period between February 2014 and August 2014 ({\it right}).
  The light curves are vertically shifted for clarity.}
\label{lc_opt}
\end{figure*}

The data reduction was standardized, including bias, dark current, flatfield, astrometry and astrometric distortion corrections 
performed with IRAF\footnote{IRAF is distributed by the National Optical Astronomy Observatory, which is operated by the Association of Universities for Research 
in Astronomy (AURA) under cooperative agreement with the National Science Foundation.} in combination with SCAMP (\citealt{2006ASPC..351..112B}) and SWARP 
(\citealt{2002ASPC..281..228B}) routines. A 7$\farcs$5 diameter aperture was used to extract the flux, which was normalized and converted into photometric fluxes through the comparison with 17 non-
variable stars located on the same images within 30$\arcmin$ around the AGN and of similar brightness as the AGN. Absolute calibration was performed using 
standard reference stars from \cite{2009AJ....137.4186L} observed on the same nights as the AGN, considering the atmospheric (\citealt{2011A&A...527A..91P}) and 
galactic foreground extinction (\citealt{2011ApJ...737..103S}) corrections. The reader is referred to \cite{2013A&A...552A...1P} for further information about 
reduction and light curve extraction process.\\\\
BEST-II\\\\
Broad-band $B$ ($4330$\,\AA) and $R$ ($7000$\,\AA) Johnson observations were carried out during a monitoring campaign between May 01 and July 30 of 2013, and 
between February 03 and August 19 of 2014, with the robotic 25\,cm Berlin Exoplanet Search Telescope-II (BEST-II). The BEST-II telescope is equipped with a Peltier-
cooled $4096 \times 4096$ pixel Finger Lakes Imager CCD KAF-16801, yielding a field-of-view of $1.7^{\circ} \times 1.7^{\circ}$ with a pixel size of 9 $\mu m$. 
More information about the telescope and the instrument has been published by \cite{2009A&A...506..569K}. The data were reduced and calibrated following the same 
procedures as for the VYSOS-6 telescope.\\\\
BMT\\\\
Narrow-band ($6700\pm 120$\,\AA\,, hereafter denoted as 670) $Ashahi$ observations were carried out simultaneously to the broad-band monitoring campaign in 2014, using 
the robotic 40\,cm Bochum Monitoring Telescope (BMT). The BMT telescope is equipped with a two-stage thermoelectric cooled $3072 \times 2038$ pixel CCD SBIG STL-6303, 
yielding a field-of-view of $41\arcmin$ $\times$ $27\arcmin$ with a pixel size of 9 $\mu m$. The data were reduced and calibrated following the same procedures as in 
the broad-band analysis. More information about the telescope and the instrument has been published by \cite{2013AN....334.1115R}.\\\\
SALT\\\\
The optical spectrum of PGC50427 was observed using the 11\,m Southern African Large Telescope (SALT) on May 10 2013 (Proposal Code: 2013-1-HETGU-001). 
The observations were performed with the Robert Stobie Spectrograph (RSS) mounted at the prime focus of the SALT. 
We used the PG0900 grating with a resolving power of $1065$ at $6050$\,\AA, and the spectrum covers a wavelength range between $3200-9000$\,\AA. The spectrum was taken 
in two identical consecutive exposures of 10 min each through the $2\arcsec \times 8\arcmin$ longslit PL 0200N001 at a parallactic angle. The detector was 
operated in normal readout mode with a $2 \times 2$ binning. A Xe spectrum was obtained after the object exposure for wavelength calibration. For flux calibration we observed the standard star G24-9. After the bias subtracted file provided by the SALT pipeline, we used standard IRAF routines for flat-field correction, cosmic ray rejection, 2D-
wavelength calibration, night sky subtraction and flux calibration. For the 1-D spectrum, we combined 7 columns, 
corresponding to $1.7738\arcsec$ ($0.1267\arcsec$ per unbinned pixel). The reduced spectrum is shown in Fig~\ref{spectrum_nbfilter}.

\subsection{Near-infrared monitoring} \leavevmode \\

IRIS\\\\
Near-infrared (NIR) $J$ (1.25 $\mu m$) and $K{s}$ (2.15 $\mu m$, hereafter denoted as K) observations were carried out between May 20 and August 11 of 2013 using 
the 0.8\,m Infrared Imaging System (IRIS) telescope. IRIS is equipped with a HAWAII-1 nitrogen-cooled detector array with $1024 \times 1024$ pixels, yielding a 
field-of-view of $12.5\arcmin$ $\times$ $12.5\arcmin$ and a resolution of 0.74$\farcs$/pixel (\citealt{2010SPIE.7735E..44H}). Images were obtained by combining 
double cycles of 20 seconds exposure time acquired with the observing sequence object-sky-object. The images were reduced using IRAF routines. Because the sky 
background emission contribution is one of the most difficult step in NIR data reduction, the sky frames observed close to the AGN were subtracted from each 
science frame before flat-field and further corrections. One final image, resulting from the combination of all individual frames, is obtained in order to remove 
cosmic rays, hot pixeles and negative residuals from the sky-subtracted science frames. The data reduction steps after the sky background subtraction and correction for cosmic rays, hot pixels and negative residuals, are the same as outlined for the VYSOS-6 telescope, including astrometry and astrometric distortion correction with SCAMP and SWARP. Light curves were calculated relative to 6 non-variable stars located on the same field 
having similar brightness as the AGN. Photometric calibration was achieved by using 4 high-quality flag (AAA) Two Micron All Sky Survey (2MASS) stars appearing in the same field as 
the AGN. As already noted for the optical treatment of the data (\citealt{2013A&A...552A...1P}), analysis for different aperture 
photometry was performed considering the proper minimization of the host-galaxy contribution and a 7$\farcs$0 diameter aperture was chosen for 
further analysis.

\section{Results and discussion}

\subsection{Optical light curves and BLR geometry}

The optical light curves for campaigns 2011 and 2014 are shown in Fig~\ref{lc_opt}. The light curves are published at the CDS. In the campaign of 2011, the $B$-band shows a gradual 
flux increase by 20\% from the beginning of March until a maximum is reached at the beginning of April (Fig.3, left). After this overall maximum, the fluxes 
undergo an abrupt drop by about 40\% until the end of May. Between the beginning of June and the end of August 2011, the fluxes are steady. The $r$-band light curve follow the same features than the $B$-band light curve albeit with a smaller variability amplitude. The latter is expected due to the larger constant host-galaxy contribution. Moreover, the $r$-band also contains a contribution from the strong H$\alpha$ emission line 
(Fig~\ref{spectrum_nbfilter}). The narrow SII-band light curve follows the continuum dominated broad-band light curves qualitatively, but we can see a 20 day delay compared to the continuum variability at the minimum observed at the end of June.

\begin{figure}
  \centering
  \includegraphics[width=\columnwidth]{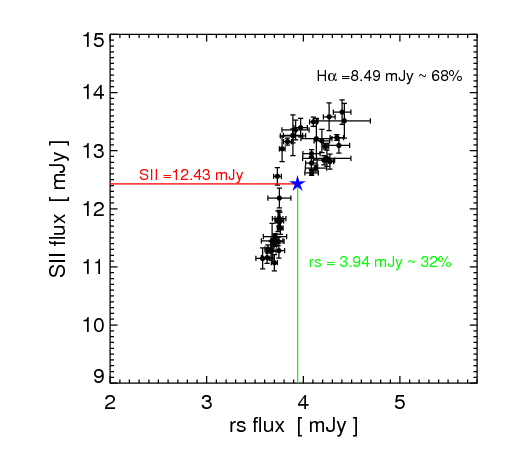}
  \caption{Flux-flux diagram for the SII and r filter. Black dots denote the measurement pair of each night during campaign 2011. The red and green lines 
  represent the average flux in the SII and r band respectively. Fluxes were measured using circular 7.5" apertures. The data are as observed and not corrected 
  for extinction.}
\label{fvg_line_first}
\end{figure}

\begin{figure}
  \centering
  \includegraphics[width=\columnwidth]{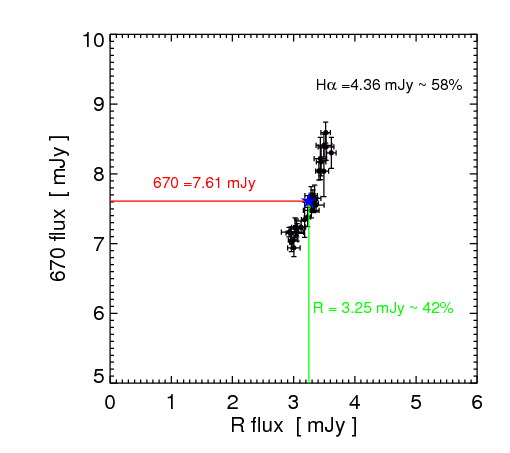}
  \caption{Same as Fig~\ref{fvg_line_first}, but for 670 and $R$ data obtained during campaign 2014.}
\label{fvg_line}
\end{figure}

\begin{figure}
  \centering
  \includegraphics[width=\columnwidth]{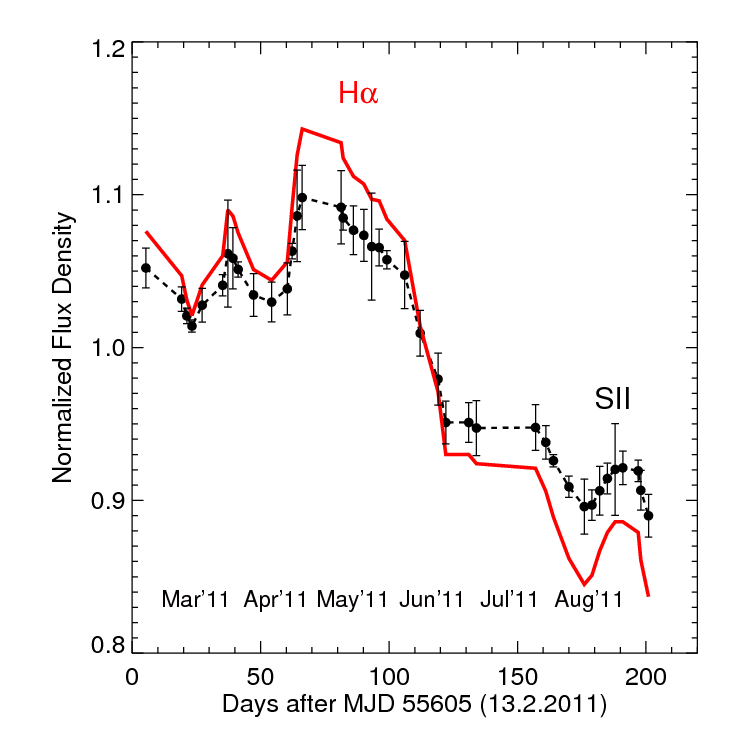}
  \caption{H$\alpha$ light curve (red) obtained after the continuum subtraction process performed on the SII light curve (black) during campaign 2011. The 
  continuum fraction was obtained from the flux-flux diagnostic on the SII and r bands; see text for details.}
\label{lc_pure_2011}
\end{figure}

\begin{figure}
  \centering
  \includegraphics[width=\columnwidth]{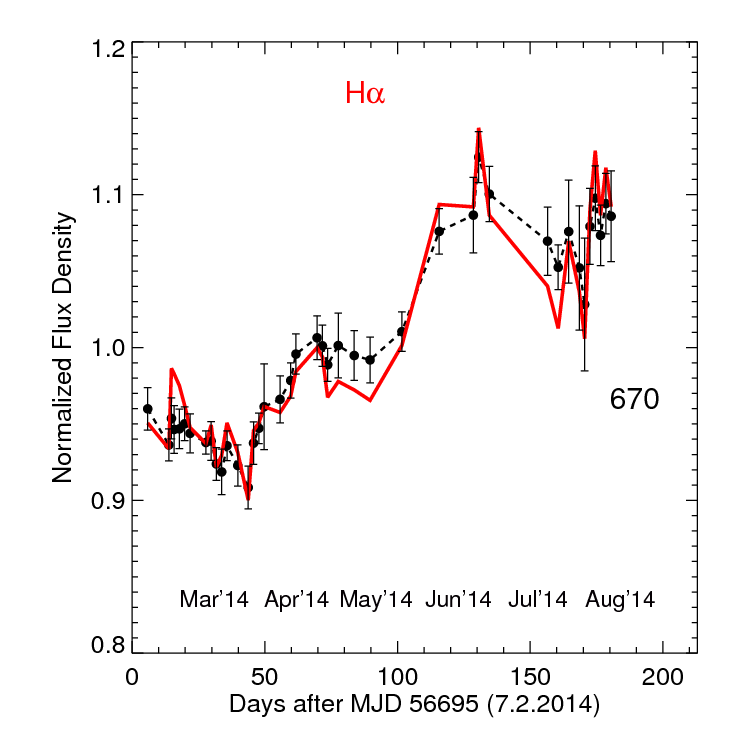}
  \caption{Same as Fig~\ref{lc_pure_2011}, but for 670 and $R$ data obtained during campaign 2014.}
\label{lc_pure_2014}
\end{figure}

As already discussed in previous PRM studies, the narrow-band contains, in addition to the H$\alpha$ line, a contribution of the varying AGN continuum, 
which must be removed before applying cross correlation techniques (\citealt{2011A&A...535A..73H}; \citealt{2012A&A...545A..84P}; \citealt{2013A&A...552A...1P}). 
In order to determine this contribution, we used the SII and $r$-band fluxes, previously calibrated to mJy, as is shown with the flux-flux diagram 
in Figure 4. The H$\alpha$ line is contributing, on average, about 70\%  of the total flux enclose in 
the SII-band, while the continuum contribution ($r$-band) is about 30\%. Following the usual practice of narrow-band PRM, we construct a 
synthetic H$\alpha$ light curve by subtracting a fraction of the $r$-band light curve (H$\alpha$ = SII $-$ 0.3 $r$), as illustred in Fig~\ref{lc_pure_2011}. The 
H$\alpha$ light curve was used afterwards to estimate
the time delay. For this purpose, we used the discrete correlation function (DCF, \citealt{1988ApJ...333..646E}) to cross-correlate the continuum 
and the synthetic H$\alpha$ emission line, taking into account possible bin size dependency\footnote{We note that the choice of a lower or higher time-bin size does not change the results if a well sampled 
data is used (\citealt{2012A&A...545A..84P}).} (\citealt{1989A&A...219..101R}).

The centroid of the cross-correlation between the $B$-band and H$\alpha$ light curves shows a time 
delay of 20.5 days (Fig 8). Uncertainties in the time delay were calculated using the flux randomization and random subset selection method (FR/RSS, 
\citealt{2004ApJ...613..682P}). From the observed light curves we create 2000 randomly selected subset 
light curves, each containing 63\% of the original data points (the other fraction of points are unselected 
according to Poisson probability). The flux value of each data point was randomly altered consistent with its (normal-distributed) measurement error.
We calculated the DCF for the 2000 pairs of subset light curves and the corresponding centroid. From this cross-correlation 
error analysis, we measure a median lag of $\tau_{cent}$ = 20.4 $^{+0.4}_{-1.0}$ $B$/H$\alpha$. Correcting for the time dilation factor ($1+z=1.0236$) we obtain a rest frame lag of $19.9 \pm 0.68$ days $B$/H$\alpha$.

\begin{figure}
  \centering
  \includegraphics[width=\columnwidth]{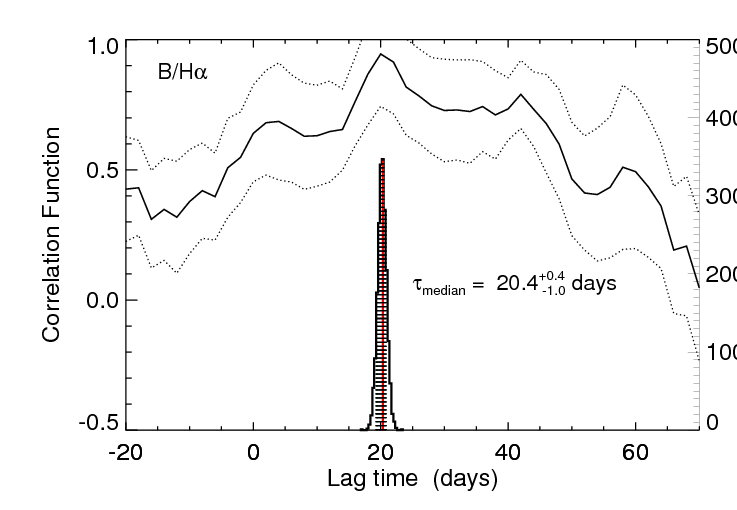}
  \caption{Cross correlation of $B$ and H$\alpha$ light curves for campaign 2011. The dotted lines indicate the error range ($  \pm
    1\sigma$) around the cross correlation. The centroid was calculated above the correlation level at $r \geq 0.8r_{max}$. The histogram shows   the distribution of the centroid lag obtained by cross correlating 2000 flux randomized and randomly selected subset light curves 
   (FR/RSS method). The black shaded area marks the 68\% confidence range used to calculate the errors of the centroid.}
  \label{CC_B_halpha_2011}
\end{figure}

\begin{figure}
  \centering
  \includegraphics[width=\columnwidth]{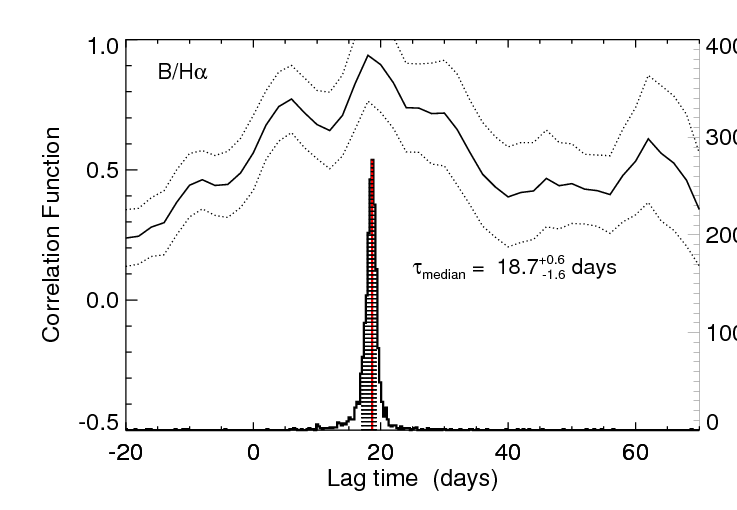}
  \caption{Same as Fig~\ref{CC_B_halpha_2011}, but for campaign 2014.}
  \label{CC_B_halpha_2014}
\end{figure}

In the campaign of 2014, we see a steep $B$-band rise at the beginning of April by about 20\% which is also observed in the $R$-band but with a smaller amplitude 
(Fig.3, right). In contrast to the steep $B$-band flux increase, the narrow 670-band rise appears stretched until the end of April with an amplitude of about 
13\%, suggesting a time delay of the H$\alpha$ line of 15-20 days. After a period of constant flux, the $B$ and $R$ light curves show a small sharp bump in 
June of about 10\%, which is observed about 15 days later in the 670-band. However, due to insufficient data obtained at this period in the 670-band, the delay cannot be measured as precise as for the first. We have isolated the H$\alpha$ emission line by the subtraction of the underlying continuum following a similar 
procedure as for the 2011 campaign. The H$\alpha$ line is contributing, on average, about 60\% of the total flux enclosed in the 670-band, while the continuum 
contribution ($R$-band) is about 40\% (Fig 5). We construct a synthetic H$\alpha$ light curve by subtracting a fraction of the $R$-band light curve (H$\alpha$ = 
670 $-$ 0.4 $R$), as illustred in Fig 7. The centroid from the cross-correlation between the $B$-band and H$\alpha$ shows a time delay of 19.3 days (Fig 9). From the FR/RSS 
cross-correlation error analysis, we measure a median lag of $\tau_{cent}$ = 18.7 $^{+0.6}_{-1.5}$. Correcting for the time dilation factor we obtain a rest 
frame lag of $18.3 \pm 1.03$ days. The deduced values for the time delay obtained at two different epochs are in qualitative agreement, and consistent with the small luminosity variations in 
the AGN ($\sim 10\%$) accross the monitoring campaign (see Section 3.3). We note that the relatively prominent contribution of the H$\gamma$ emission line (Fig. 2) to the $B$-band ($\sim 10\%$) 
may result in a slightly higher time lag centroid from the DCF as the $B$-band light curve contains not only the AGN continuum but also the lagged H$\gamma$ signal. In the net effect, our estimate $B$/H$\alpha$ may underestimate the true $B$/H$\alpha$ time delay. However, the current data does not allow to quantify this effect; this would requiere the use of emission line free continuum bands, for instance with the ultraviolet band.

\begin{figure}
  \centering
  \includegraphics[width=\columnwidth]{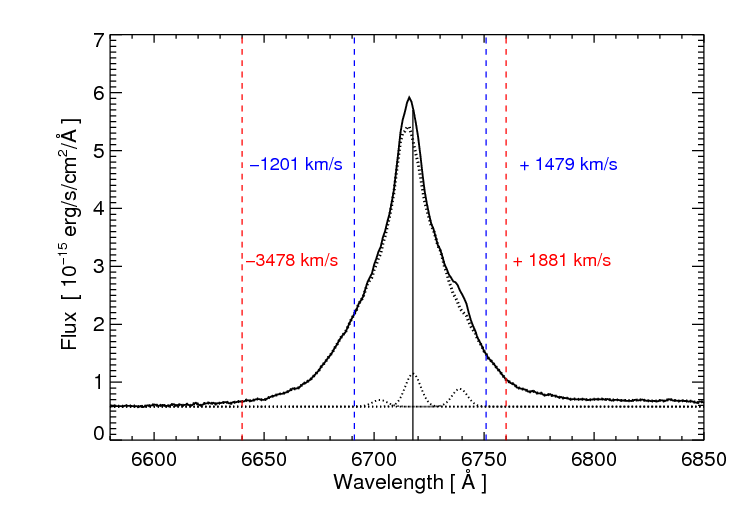}
  \caption{SALT spectrum of \object{PGC50427}, zoomed onto the H$\alpha$ line.
      The dotted black curve represents the spectrum after
      subtracting the narrow [NII]$\lambda\lambda$6548,6583 emission
      lines. The narrow H$\alpha$ and [NII] models are shown in dotted lines at the bottom of the H$\alpha$ profile.
      The NB 670 and SII filters efectively covers the line between velocities -3478km/s and +1881km/s (red dotted lines) and -1201km/s and +1479km/s (blue dotted lines) respectively.
  } 
\label{narrow_subtracted}
\end{figure}

\subsection{Central black hole mass}

Assuming that the BLR emiting gas clouds are in virialized motion around the central black hole, the mass of the black hole can be 
estimated as $M_{\rm BH} = f \cdot R_{\rm BLR} \cdot \sigma_{V}^2 / G$. The velocity $\sigma_{V}$ of the emission-line region is determined from the line dispersion ($\sigma_{\rm line}$) or from the full 
width at half maximum (FWHM) of the line profile. The scaling factor $f$ depends on the geometry and kinematics of the BLR. Most of the results presented in previous reverberation studies have been carried 
out considering only the virial product $c \tau \sigma_{V}^2/G$, ie. assuming a scaling factor $f=1$ (Peterson et al. 2004, and references therein).

The broad H$\alpha$ emission line is blended with the narrow H$\alpha$ and [NII]$\lambda\lambda$6548,6583 narrow emission lines. In order to remove the narrow 
components, we model the [SII]$\lambda\lambda$6716,6731 doublet with a multi-Gaussian profile as described in \cite{2004ApJ...610..722G}. The model is shifted and 
scaled to fit the H$\alpha$ + [NII]$\lambda\lambda$6548,6583 narrow lines and subtracted from the observed broad H$\alpha$ line profile as shown in Fig. 10. The ratio of the [NII] lines is 
fixed at the theoretical value of 2.96 and the relative positions of the narrow H$\alpha$ and [NII] lines are determined by their laboratory wavelenghts. After 
removing the narrow emission lines, the H$\alpha$ profile was isolated by the subtraction of a linear continuum fit, obtained through 
interpolation between two continuum segments on either end of the line. Figure 10 illustrate the original H$\alpha$ emission line profile together with the subtracted narrow 
emission line profiles. The velocity dispersion after removal of narrow lines is $\sigma_{\rm line} = 1020$ km s$^{-1}$, which has been corrected for instrumental velocity dispersion to obtain an intrisic profile width. The feasibility of the use of single-epoch (SE) spectra for the black
hole mass determination has been established in previous investigations. On average, uncertainties of $\sim$30\% have been reported for black hole mass determination from single epoch spectra measurements (e.g. \citealt{2002ApJ...571..733V}, \citealt{2007ApJ...661...60W}, \citealt{2009ApJ...692..246D}).

\begin{figure}
  \centering
  \includegraphics[width=\columnwidth]{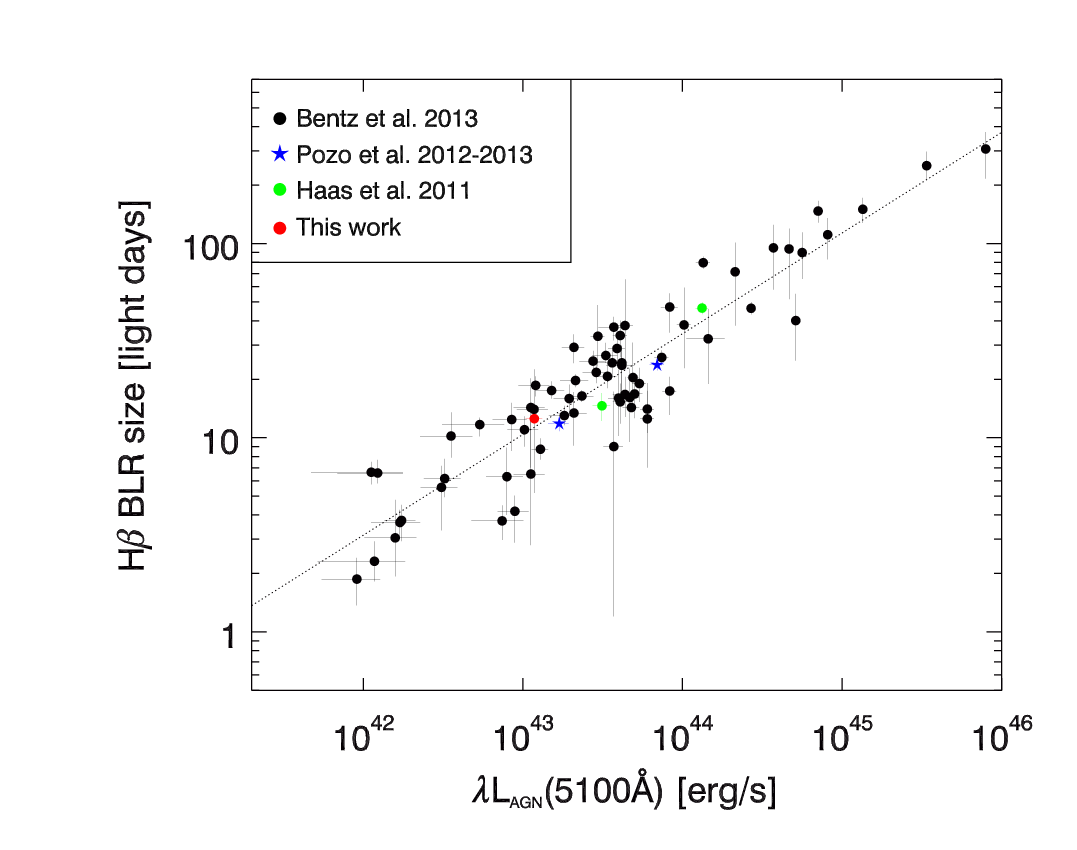}
  \caption{$R_{BLR} - L$ relationship from data of \cite{2013ApJ...767..149B} (black dots)
    with a fitted slope $\alpha = 0.533$ (dotted line). The diagram also contain the objects from previous photometric reverberation mapping campaigns 
    (\citealt{2011A&A...535A..73H}, green dots; \citealt{2012A&A...545A..84P}-2013, blue stars, and this article, red filled circle).}
\label{blr_lum}
\end{figure}

\begin{figure*}
  \centering
  \includegraphics[width=0.67\columnwidth]{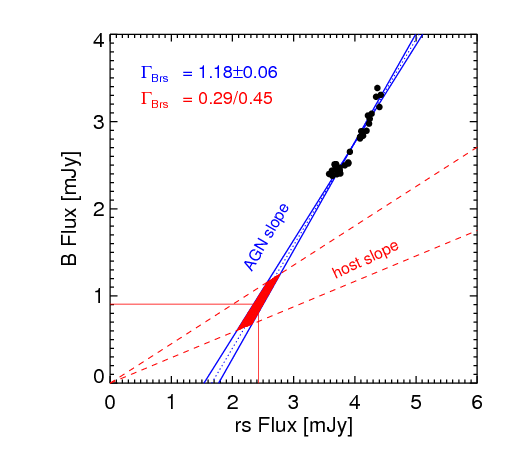}
  \includegraphics[width=0.67\columnwidth]{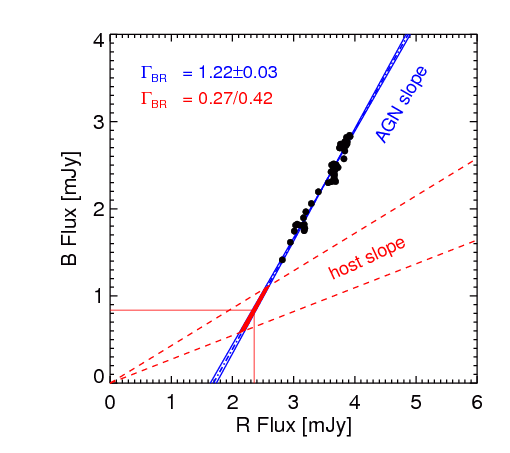}
  \includegraphics[width=0.67\columnwidth]{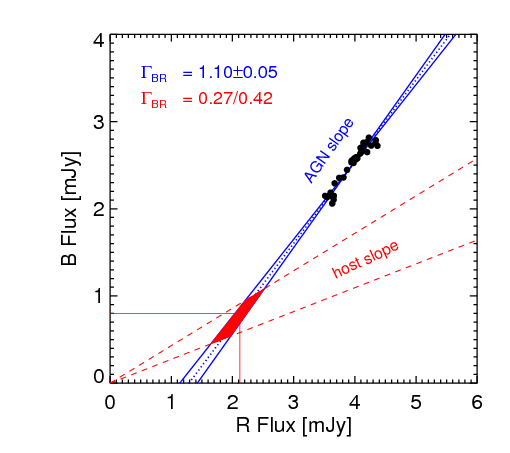}
  \caption{Flux variation gradient diagrams for 2011 ({\it left}), 2013 ({\it middle}) and 2014 ({\it right}) campaigns. 
    The solid lines delineate the
    bisector regression model yielding the range of the AGN slope. 
    The dashed lines indicate the range of host slopes determined
    by \cite{2010ApJ...711..461S} for 11 nearby AGN. 
    The intersection between the host galaxy and AGN slope (red area)
    gives the host galaxy flux at the time of the campaign in both bands.}
  \label{fvg}
\end{figure*}

Using the derived time delay $\tau = 18.3$d for epoch 2014 and the velocity dispersion (with 30\%
uncertainty), the virial black hole mass is $M_{virial} = (3 \pm 2) \times
10^{6} M_{\odot}$. Considering the factor $f=5.5 \pm 1.8$, based on the asumption that AGNs follow the same $M_{\rm BH} - \sigma_{*}$ relationship (Onken et al. 2004), we determine a central black hole mass 
$M_{BH} =(17 \pm 11) \times 10^{6} M_{\odot}$. 

Assuming a symmetric BLR, the dimensionless factor $f$ depends on the unknown inclination of the BLR ($f=\frac{2 \cdot \ln2}{\sin^2{i}}$). The optical and the H$\alpha$ emission line variability observed in PGC50427 suggest a disk-
like BLR geometry with low inclination $i\le30^\circ$ (Pozo Nunez et al., in preparation), hence the geometry-scaling factor $f$ may be much higher than the commonly used $f=5.5$. Therefore the black hole 
mass derived here should be considered as lower limit.

\subsection{Host-subtracted nuclear luminosity and the BLR size -- luminosity relationship}

To determine the AGN luminosity free of host galaxy contributions, we
applied the flux variation gradient  
(FVG) method, originally proposed by \cite{1981AcA....31..293C} and later
modified by \cite{1992MNRAS.257..659W}.  
A detailed description of the FVG method on PRM data is presented in
\cite{2012A&A...545A..84P}. In this method the fluxes obtained through different filters and same
apertures are plotted in a flux-flux diagram. The fluxes follow a
linear slope representing the AGN color, while the slope of the
nuclear host galaxy contribution 
(including the contribution from the narrow line region (NLR)) lies in
a well defined range  ($0.4 < \Gamma^{host}_{BV} <
0.53$, for 8$\farcs$3 aperture and redshift $z<0.03$, \citealt{2010ApJ...711..461S}). 
The AGN slope is determined through a linear
regression analysis. Averaging over the intersection area between the AGN and the host galaxy slopes
yields the actual host galaxy contribution at the time of
the monitoring campaign. Figure~\ref{fvg} shows the FVG
diagram for the $B$ and $r$ fluxes corresponding to campaign 2011 and for the $B$ and $R$ fluxes corresponding
to campaign 2013/2014 obtained during the same nights and through a $7\farcs5$
aperture. All the fluxes has been corrected for galactic foreground extinction.

The bisector linear regression method yields a linear gradient of $\Gamma_{BR_{s}} = 1.18 \pm 0.06$, $\Gamma_{BR} = 1.22 \pm 0.03$ and  
$\Gamma_{BR} = 1.10 \pm 0.05$ for campaigns in 2011, 2013 and 2014 respectively. The results are consistent, within the uncertainties, with the
gradients obtained for other Seyfert 1 galaxies by \cite{1992MNRAS.257..659W} and \cite{2010ApJ...711..461S}. The AGN fluxes at the time of the monitoring can be 
determined by subtracting the host galaxy fluxes from the total fluxes. The host galaxy subtracted average AGN fluxes and
the host galaxy flux contribution of \object{PGC50427} are listed in
Table~\ref{table4}. Also listed in Table~\ref{table4} are the interpolated rest frame
$5100\AA~$ fluxes and the monochromatic AGN luminosity $\lambda
L_{\rm{\lambda(AGN)}}$ at $5100\AA~$ obtained at the distance of 102 Mpc. The rest frame flux at 5100\AA~ was interpolated from the host-subtracted AGN fluxes in 
both bands, assuming for the interpolation that the AGN spectral energy distribution (SED) is a power law ($F_{\nu} \propto
\nu^{\alpha}$) with a spectral index $\alpha=\log(fx_{AGN} / fy_{AGN}) / \log(\nu_{x} / \nu_{y})$,
where $\nu_{x}$ and $\nu_{y}$ are the effective frequencies in the X and Y bands, respectively.
The error was determined by interpolation between the ranges of the AGN fluxes $\pm\sigma$ 
in both filters, respectively.

The position of PGC50427 on the BLR size-luminosity diagram is shown in Figure~\ref{blr_lum}. The values of other galaxies are taken from \cite{2013ApJ...767..149B} and
from previous photometric reverberation mapping campaigns (\citealt{2011A&A...535A..73H}, \citealt{2012A&A...545A..84P}, 2013). For this figure we converted the measured H$\alpha$ size into the size of the H$\beta$ BLR using the weighted mean ratio for the time lag $\tau(H\alpha):\tau(H\beta):1.54:1.00$, obtained by Bentz et al. (2010) from the Lick AGN Monitoring Program of 11 low-luminosity AGN.

\begin{table*}
\begin{center}
\caption{Total, host galaxy and AGN optical fluxes given in mJy for the different campaigns.}
\label{table4}
\begin{tabular}{@{}ccccccc}
\hline\hline
Campaign & Filter & Total & Host & AGN$^{1}$ & $f_{AGN} ((1+z)5100$\AA~ & $\lambda L_{\lambda,AGN}5100$\AA~\\
         &        & (mJy) & (mJy) & (mJy) & (mJy) & ($10^{43} erg s^{-1}$) \\   
\hline
2011 & $B$ & 2.71$\pm$0.12 & 0.91$\pm$0.17 & 1.80$\pm$0.20 & 1.66$\pm$0.20 & 1.18$\pm$0.14 \\
     & $rs$ & 3.95$\pm$0.11 & 2.42$\pm$0.15 & 1.53$\pm$0.19 \\
2013 & $B$ & 2.36$\pm$0.10 & 0.83$\pm$0.15 & 1.53$\pm$0.18 & 1.40$\pm$0.17 & 1.00$\pm$0.12 \\
     & $R$ & 3.56$\pm$0.11 & 2.35$\pm$0.12 & 1.21$\pm$0.16 \\
2014 & $B$ & 2.48$\pm$0.10 & 0.80$\pm$0.13 & 1.68$\pm$0.16 & 1.72$\pm$0.17 & 1.22$\pm$0.12  \\
     & $R$ & 3.94$\pm$0.12 & 2.12$\pm$0.15 & 1.82$\pm$0.19 &  &  \\          
\hline
\end{tabular}
\end{center}
\tablefoottext{1}{AGN fluxes values $f_{AGN} = f_{total}-f_{host}$ with uncertainty range $\sigma_{AGN} = (\sigma_{total}^{2} + \sigma_{host}^{2})^{0.5}$.}\\  
\end{table*}

\begin{figure}
  \centering
  \includegraphics[width=\columnwidth,clip=true]{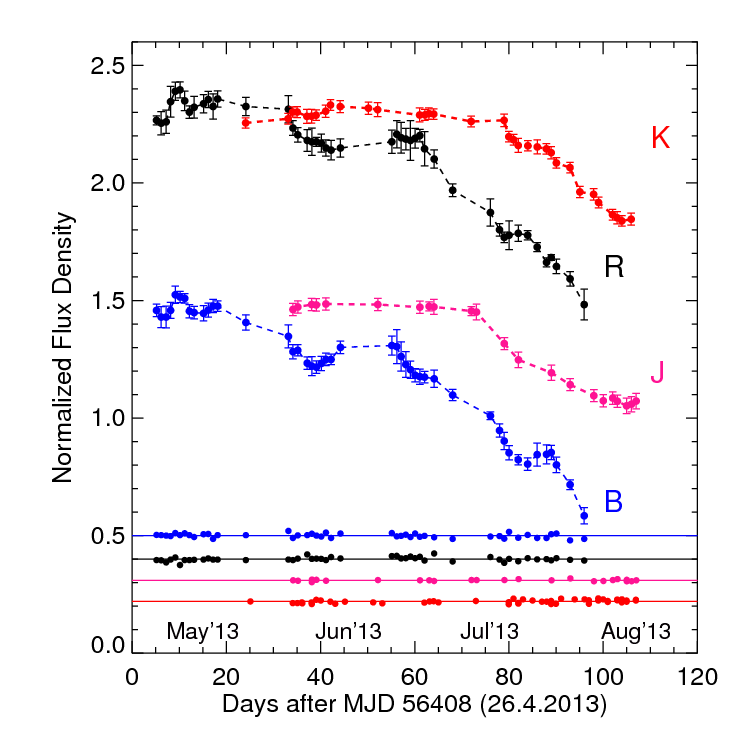}
  \caption{Optical and NIR host galaxy subtracted light curves obtained between May and August 2013.}
  \label{optical_inf}
\end{figure}

\subsection{Infrared light curves and Dust-torus size}

\begin{figure}
  \centering
  \includegraphics[width=\columnwidth]{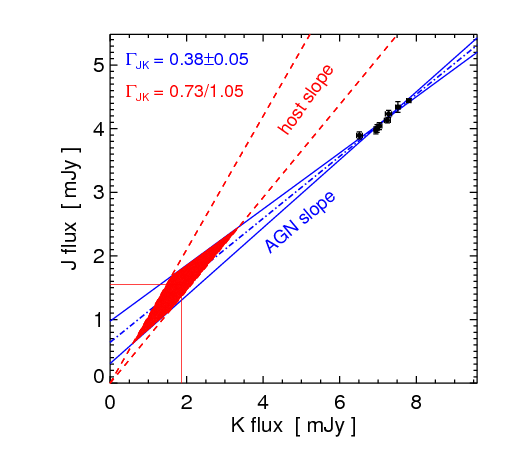}
  \caption{Flux variation gradient diagram in the NIR.
    The data are represented by the black dots.
    The solid blue lines represent the
    bisector fit, yielding the range of the AGN slopes. 
    The dashed red lines indicate the range of host slopes determined
    by \cite{2006ApJ...639...46S}. The intersection between the host galaxy and AGN slope (red area)
    gives the host galaxy flux in the respective bands.}
\label{fvg_inf}
\end{figure}

\begin{figure}
  \centering
  \includegraphics[width=\columnwidth,clip=true]{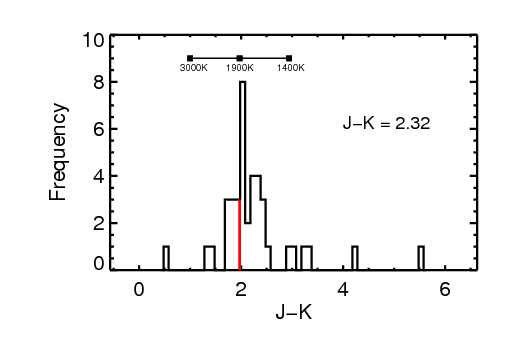}
  \caption{Distribution of AGN NIR $J-K$ colors of Seyfert galaxies taken from \citealt{2004MNRAS.350.1049G} after host galaxy subtraction. The red vertical solid line
  correspond to the $J-K$ color of PGC50427. The filled black squares indicate Blackbody color temperatures.}
  \label{glass_plot}
\end{figure}

Figure~\ref{optical_inf} depicts the optical and near-infrared normalized light curves of the nucleus of PGC50427 obtained during the 2013 campaign. The light curves are published at the CDS. To deconvolve
the host galaxy and the nuclear flux contributions, we used the flux variation gradient (FVG) method in the same way as described in \cite{2014A&A...561L...8P}. The spectral energy distribution of the variable component remain constant with time, and in consequence the slopes obtained from the optical and NIR flux ratios (B/R and J/K) allow us to separate the AGN flux through the use of a well-defined range
of host galaxy colors (\citealt{2004MNRAS.350.1049G}, \citealt{2010ApJ...711..461S}). Figure~\ref{fvg_inf} shows the $J$ flux plotted 
against the $K$ flux for each night. The bisector linear regresion method yields a linear gradient of $\Gamma_{JK} = 0.38 \pm 0.05$ which correspond to a nuclear color $J-K = 1.97 \pm 0.07$, and is consistent 
with the FVG colors of Seyfert galaxies ($J-K \sim 2.3$) measured by \cite{2004MNRAS.350.1049G}. Figure~\ref{glass_plot} shows the distribution of $J-K$ colors of Seyfert 
galaxies previously studied by \cite{2004MNRAS.350.1049G} and the value for PGC50427 obtained in this work. The color of PGC50427 correspond to a blackbody temperature of 1850K, which is in agreement with the sublimation 
temperature of hot pure graphite grains (\citealt{2012MNRAS.420..526M}), and supports the previous conclusions that IR emission observed from Seyfert galaxies is dominated by thermal radiation from the hot dust 
close to the central AD (\citealt{2006ApJ...639...46S}).

The optical $B$ and $R$-bands, which are mostly dominated by the AGN continuum\footnote{We note that the $R$-band also contains a small contribution from the strong H$\alpha$ emission line.}, show a similar 
varibility behavior. The NIR $J$ and $K$-bands, which are dominated by the innermost hot dust reacts with delayed response to the AGN continuum.

\begin{figure}
  \centering
  \includegraphics[width=\columnwidth]{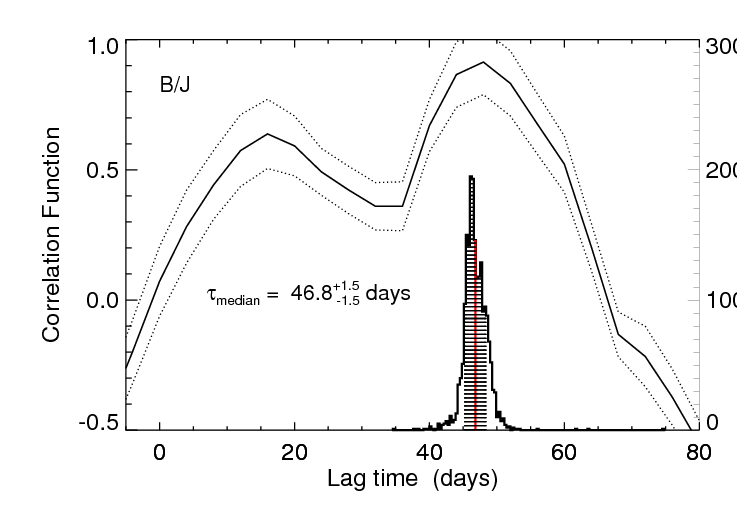}
  \caption{Cross correlation of $B$ and $J$ light curves. The dotted lines indicate the error range ($  \pm
    1\sigma$) around the cross correlation. The centroid was calculated above the correlation level at $r \geq 0.6r_{max}$. The histogram shows the
    distribution of the centroid lag obtained by cross correlating 2000 flux randomized and randomly selected subset light curves 
    (FR/RSS method). The black shaded area marks the 68\% confidence range used to calculate the errors of the centroid.}
  \label{dcf_inf_first}
\end{figure}

\begin{figure}
  \centering
  \includegraphics[width=\columnwidth]{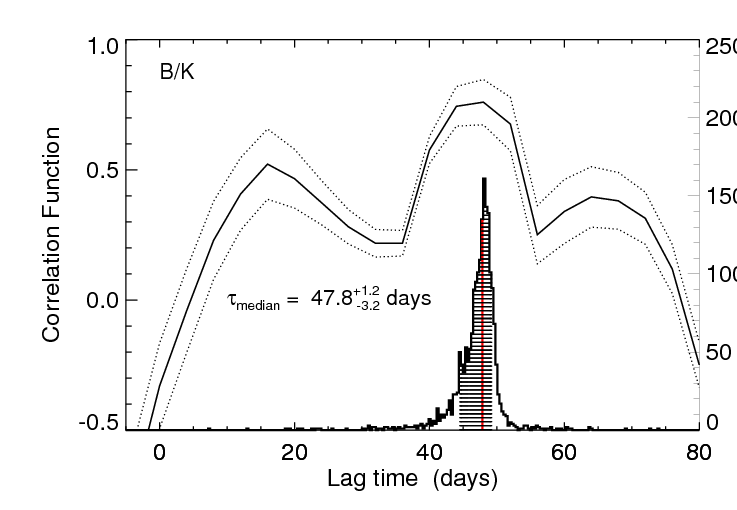}
  \caption{Same as Fig~\ref{dcf_inf_first}, but for $B$ and $K$ light curves.}
  \label{dcf_inf_second}
\end{figure}

The time delay between the AGN continuum and dust emission 
can be estimated by cross-correlation of the optical and NIR light curves yielding the average radius of the innermost dust torus. We correlated both the $B$- and $K$-band light curves and the $B$- and $J$-
band light curves using the discrete correlation function (DCF, \citealt{1988ApJ...333..646E}). The cross correlation of $B/J$ shows two peaks, one small correlation peak around lag 19 days obtained above the 
correlation level at $r \geq 0.4r_{max}$ and a major peak with a lag of 46 days obtained above the correlation level at $r \geq 0.6r_{max}$, as shown in Figure~\ref{dcf_inf_first}. Similar features can be seen 
in the cross correlation of $B/K$, a small correlation peak with a lag of 19 days, a major peak with a lag of 47 days, and an additional third and smallest peak with a lag of 67 days as shown in 
Figure~\ref{dcf_inf_second}. The NIR host galaxy corrected light curves show features of similar amplitude and sharpness as the optical host galaxy corrected light curves. In addition, the cross correlation are at nearly zero level at lag 0 days. One expects that if the dust distribution is spread at different line-of-sight distances, the observed echo will be smeared out in time and the $J$ and $K$ light curves will show a smoother variability. Therefore the observed NIR light curves of PGC50427 argues in favor of a face-on torus geometry.    

The uncertainty of the lag time $\tau$ was estimated using the flux randomization and random subset-selection method (FR and RSS, \citealt{2004ApJ...613..682P}). The median of this procedure yields 
$\tau_{cent}$ = 46.8 $^{+1.5}_{-1.5}$ days and $\tau_{cent}$ = 47.8 $^{+1.2}_{-3.2}$ days for $B/J$ and $B/K$, respectively. Correcting for the time dilation factor, we obtain a rest frame time-delay 
$\tau_{rest} = 45.7 \pm 1.47$ days and $\tau_{rest} = 46.7 \pm 2.15$ days for $B/J$ and $B/K$, respectively. 

\begin{figure}
  \centering
  \includegraphics[width=\columnwidth,clip=true]{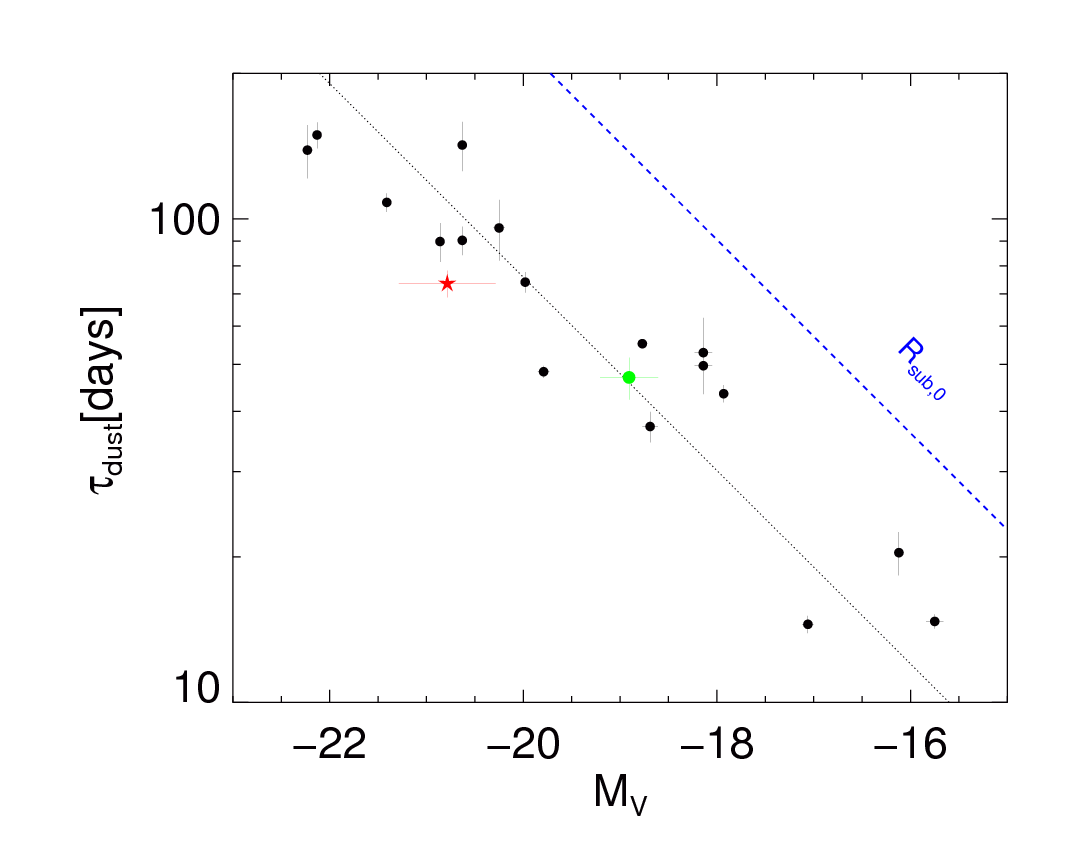}
  \caption{Lag -- luminosity relationship based on the data of \cite{2006ApJ...639...46S} and \cite{2014ApJ...788..159K}. The solid line is the best-fit 
  regression from \cite{2014ApJ...788..159K}. PGC50427 (green dot) lies
    close to the regression line. The dashed blue line indicates the dust
    sublimation radius $r_{sub}$ expected at a given
    nuclear luminosity $M_{V}$ (from \citealt{2007A&A...476..713K}).}
  \label{torus_lum}
\end{figure}

Dust-reverberation studies of Seyfert 1 galaxies have shown that the dust torus size obtained by the cross-correlation of the optical $V$ and $K$-bands is proportional to the square
root of the optical luminosity $\tau \propto L^{0.5}$ (Suganuma et al. 2006, Koshida et al. 2014). Figure~\ref{torus_lum} shows the position of PGC50427 on the Dust size-luminosity diagram. To obtain the optical $V$-band fluxes we interpolated the AGN $B$ and $R$ band fluxes.

\section{Summary and conclusions}
\label{section_conclusions}

We have performed three monitoring campaigns between 2011 and 2014, as well as obtaining a SALT spectrum, of the Seyfert 1 galaxy \object{PGC50427}. We determined the broad line region
size, the dust torus size, the virial black hole mass and the host-subtracted AGN optical and NIR
luminosity. The results are:

\begin{enumerate}

\item The cross-correlation of the H$\alpha$ emission line with the optical continuum during campaign 2011 yields a rest-frame
time delay $\tau_{rest}$ = 19.9 $\pm 0.68$ days. During campaign 2014 the cross-correlation
of the H$\alpha$ emission line with the optical continuum yields a rest-frame time delay $\tau_{rest}$ = 18.3 $\pm 1.03$.

\item With the velocity dispersion obtained from a single epoch spectrum in 2013, we determine the black hole mass $M_{BH} =(17 \pm 11) \times 10^{6} M_{\odot}$ assuming a geometrical factor of 5.5.  

\item Using the flux variation gradient method (FVG) we determine the host galaxy subtracted optical AGN luminosity of PGC50427 at different epochs
of observations  $L_{\rm{AGN-2011}} = (1.18 \pm 0.14)\times 10^{43}erg s^{-1}$,  $L_{\rm{AGN-2013}} = (1.00 \pm 0.12)\times 10^{43}erg s^{-1}$, and
$L_{\rm{AGN}} = (1.22 \pm 0.12)\times 10^{43}erg s^{-1}$.

\item From the NIR light curves in 2013, we determine a lag time of $\tau_{rest} = 45.7 \pm 1.47$ days and $\tau_{rest} = 46.7 \pm 2.15$ days for $B/J$ and $B/K$, respectively. The relatively sharp dust echo 
observed in the NIR light curves argues in favor of a face-on torus geometry. The infrared lag time is correlated with the optical luminosity according to $R_{dust} \propto L^{0.5}$ in agreement with previous investigations. The infered inner size for the dust
torus in PGC50427 suggest that the location of the thermal emitting region is located well outside the BLR which support the unified scheme of AGNs.

\end{enumerate}

\begin{acknowledgements}
 
  This publication is supported as a project of the
  Nordrhein-Westf\"alische Akademie der Wissenschaften und der K\"unste
  in the framework of the academy program by the Federal Republic of
  Germany and the state Nordrhein-Westfalen, by Deutsche Forschungsgemeinschaft (DFG HA3555/12-1) 
  and by Deutsches Zentrum f\"ur Luft-und Raumfahrt (DLR 50\,OR\,1106). 
  
  This work has been supported by DFG grant Ko857/32-2.
  
  The observations on Cerro Armazones benefitted
  from the care of the guardians Hector Labra, Gerardo Pino, Roberto Munoz, 
  and Francisco Arraya.
  
  Construction of the IRIS infrared camera was supported by the National Science Foundation
  under grant AST07-04954.
  
  Some of the observations reported in this paper 
  were obtained with the Southern African Large Telescope (SALT).  
  
  We thank Janjenka Szillat for her support regarding the spectral reduction.
  
  This research has made use of the NASA/IPAC
  Extragalactic Database (NED) which is operated by the Jet Propulsion
  Laboratory, California Institute of Technology, under contract with
  the National Aeronautics and Space Administration. This research has made 
  use of the SIMBAD database, operated at CDS, Strasbourg, France.
  
  We thank our referee Darach Watson for helpful comments
  and careful review of the manuscript.

\end{acknowledgements}

\bibliographystyle{aa} 
\bibliography{dust_BLR}

\end{document}